\documentclass[prd, preprint, 12pt]{revtex4-1}

\usepackage{amsmath}
\usepackage{amssymb}
\usepackage{setspace}
\usepackage{graphicx}
\usepackage{bbm}
\usepackage{natbib}
\usepackage{float}
\usepackage{hyperref}
\usepackage{braket}
\usepackage{braket}
\usepackage{tensor}
\usepackage[utf8]{inputenc}
\usepackage{amsfonts}
\usepackage{fdsymbol}
\usepackage{footnote}

\begin{document}
 
 %

\begin{center}
 {  \large {\bf Proposal for a New Quantum Theory of Gravity}}



{\bf Tejinder P. Singh}

{\it Tata Institute of Fundamental Research,}
{\it Homi Bhabha Road, Mumbai 400005, India}
{\tt e-mail:tpsingh@tifr.res.in}
\end{center}
\setstretch{1.24}


\centerline{\bf ABSTRACT}
\noindent We recall a classical theory of torsion gravity with an asymmetric metric, sourced by a Nambu-Goto + Kalb-Ramond string \cite{Hammond}. We explain why this is a significant gravitational  theory, and in what sense classical general relativity is an approximation to it. We propose that a non-commutative generalisation of this theory (in the sense of Connes' non-commutative geometry and Adler's Trace Dynamics) is a `quantum theory of gravity'. The theory is in fact a classical matrix dynamics with only two fundamental constants -- the square of the Planck length and the speed of light, along with the two string tensions as parameters. The guiding symmetry principle is that the theory should be covariant under general coordinate transformations of {\it non-commuting} coordinates.The action for this non-commutative torsion gravity can be elegantly expressed as an invariant area integral, and represents an atom of 
space-time-matter. The statistical thermodynamics of a large number of such atoms yields the laws of quantum gravity and quantum field theory, at thermodynamic equilibrium. Spontaneous localisation caused by large fluctuations away from equilibrium is responsible for the emergence of classical space-time and the field equations of classical general relativity. The resolution of the quantum measurement problem by spontaneous collapse is an inevitable consequence of this process. Quantum theory and general relativity, are both seen as emergent phenomena, resulting from coarse-graining of the underlying non-commutative geometry. We explain the profound role played by entanglement in this theory: entanglement describes interaction between the atoms of space-time-matter, and indeed entanglement appears to be more fundamental than quantum theory or space-time.
We also comment on possible implications for black hole entropy and evaporation, and for cosmology. We list the intermediate mathematical analysis which remains to be done to complete this programme.
\smallskip


\noindent 

\smallskip

\noindent {\textit {Remembering Albert Einstein on his 140th birthday, we hope that this paper is a useful small step towards his extraordinary vision on the future of quantum theory.}}

\smallskip

\noindent{{\it``There is no doubt that quantum mechanics has seized hold of a beautiful element of truth and that it will be a touchstone for a future theoretical basis in that it must be deducible as a limiting case from that basis just as electrostatics is deducible from the Maxwell equations of the electromagnetic field or as thermodynamics is deducible from statistical mechanics. However, I do not believe that quantum mechanics will be the starting point in the search for this basis, just as one cannot arrive at the foundations of mechanics from thermodynamics or statistical mechanics.}'' - A. Einstein, Journal of the Franklin Institute, 221, 313 (1936).
}



\newpage
\section{Introduction and Summary} 
We have argued in various earlier works that there must exist a reformulation of quantum theory which does not refer to classical time \cite{Singh:2006, Singh:2012, Singh:2017}. This applies as much to relativistic quantum field theory as it does to non-relativistic quantum mechanics. The reasoning is that the classical space-time manifold is part of a classical geometry whose metric is determined by classical bodies, via the laws of the general theory of relativity. In the absence of classical bodies (which are a limiting case of quantum theory), the metric would undergo quantum fluctuations, and then it is a consequence of the Einstein hole argument that the point structure of the underlying manifold loses its physical meaning. And yet, one ought to be able to formulate quantum theory; hence the need for a description of the theory which does not refer to classical time.

Once we have such a reformulation, there are far reaching consequences, irrespective of what the actual mathematical structure of this reformulation is. Two of these are particularly significant. The first is that once one removes classical space-time from quantum theory, the non-classical space-time which takes its place (by which we mean any structure other than classical space-time, and from which the classical space-time is recovered) is a new feature of the theory. There must exist a mechanism which makes possible the transition from 
non-classical to classical space-time, and it appears that relativistic spontaneous localisation is one such mechanism \cite{Singh:2019}. And, as a result, the non-relativistic limit of this localisation process is just what provides a solution for the quantum measurement problem, and explains the quantum-to-classical transition. Furthermore, the non-classical space-time helps to resolve the puzzle of quantum
non-locality, and provides a better understanding of the physical meaning of the wave function. Thus it seems to us that the problem of time in quantum theory is the most serious amongst all its foundational problems, and its resolution automatically takes care  of the other problems, including the infamous quantum measurement problem.

The second far-reaching consequence is that once classical space-time is lost, we no longer have at hand the recipe of `quantising' a classical theory, so as to arrive at the corresponding quantum theory. For, in the traditional quantisation process, the classical space-time is left untouched -- the space-time in classical dynamics is very much the same space-time as in quantum dynamics. But having moved to a non-classical space-time, we no longer have a recipe which takes us from classical dynamics to a quantum dynamics in
non-classical space-time. It appears that we must write down the principles of quantum theory on a non-classical space-time ab initio, and preferably be guided by a convincing symmetry principle. We note also that in the sought for reformulation of quantum theory, the gravitational field must also be non-classical, just as the manifold is replaced by something non-classical. Evidently then, the sought-for reformulation of quantum field theory without classical space-time is a candidate for a quantum theory of gravity--one which must be arrived at by first principles, and not by quantising classical general relativity.

What clues do we have for constructing such a reformulation? Canonical quantum theory is arrived at from its classical version by elevating the Poisson brackets of canonical variables to commutation relations. However, space-time coordinates are left as such, even if commutation relations for the dynamical metric degrees of freedom are introduced, as in quantum general relativity\footnote{It should be clarified that there is as such no space-time in quantum general relativity. The dynamical degrees of freedom are the three-geometries, or, in the case of loop quantum gravity, these are the holonomies}. Having concluded above that a reformulation of quantum theory must exist in which the coordinates are no longer classical, we propose to raise the coordinates to non-commuting coordinates, and propose the following symmetry principle: The laws of gravitation should be invariant under general coordinate transformations of 
{\it non-commuting} coordinates. We will construct an appropriate non-commutative classical theory of gravitation obeying this principle, and then show that a quantum theory of gravity emerges from the statistical thermodynamics of this underlying non-commutative gravity. Relativistic spontaneous localisation will be the mechanism which takes us from quantum gravity to classical general relativity. In the following paragraphs, we summarise our argument, and explain how the theory of quantum gravity is arrived at.

The domain of non-commuting space-times is Connes' non-commutative differential geometry. We assume as given a four-dimensional Lorentzian non-commutative manifold with 
a set of coordinate operators ${\hat x}^\mu$ that do not commute with each other. Since they do not commute with each other, it is reasonable to assume that the metric is {\it asymmetric}, because the anti-symmetric part of the metric will also contribute to the definition of distance. In Connes' geometry, an infinitesimal distance $ds$ is defined from the inverse of the Dirac operator $ds = D^{-1}$, allowing the definition of the line-element $ds^2$ and a {\it symmetric} metric. In his language, the two-dimensional measure of a four-manifold, i.e., its ``area", is to be
computed as the integral $\int ds^2$, which can be   shown, in the leading order in an asymptotic expansion,  to be proportional to the Einstein-Hilbert action:
\begin{equation}
\int ds^2 =     - \frac{1}{48 \pi^2} \int R\; \sqrt{-g} \; d^4x
\label{gravityaction}
\end{equation}
where $\sqrt{-g} \; d^4x$ is the volume form, and $R$ is the scalar curvature.

In Section III we will argue that this definition of distance can be extended to the case when the metric is asymmetric. Furthermore, for reasons to be elaborated below, we will define a torsion tensor from the anti-symmetric part of the metric, and the curvature scalar will depend on the connection (as usual) and also on the torsion tensor defined from the anti-symmetric part of this metric. This anti-symmetric part will be shown to be related to spin (Section II). Spin and torsion are known to naturally couple to each other; hence the need to include torsion.  Unlike in Poincar\'e local gauge theory, torsion here is defined from the anti-symmetric part of the metric, and below we explain the reason for preference over Poincar\'e gauge gravity.  The geometric part of the action in our theory will be the area integral shown in (\ref{gravityaction}), now understood to depend on an asymmetric metric as well as on the torsion tensor defined from it. In order to have a scale for this area, we assume that there exists in the theory an area unit which is the square $L_{pl}^2$ of the Planck length. This, along with the speed of light $c$, will be the only two fundamental constants of the theory. Planck's constant $\hbar$ and Newton's gravitational constant $G$ will arise from these, in an emergent sense. 

Given that the geometric part of the action is an area integral, what should the matter part of the action be? One straightforward possibility is to simply take the matter part to be the sum over Dirac fermionic fields. This, however, is unsatisfactory for various reasons. To begin with, since there is no $\hbar$, we cannot introduce the fermionic mass $m$, because we cannot make a length scale such as Compton wavelength. Also, such a Dirac action does not suggest a way of considering  this action in terms of an ``area", and an action for matter as an area integral is preferred for aesthetic reasons given that the geometric part is also an area. Remarkably enough, we do know where to get such an area dependent action from, and that is the world sheet action of a string!
We will hence assume the matter action to be the sum of two contributions: the Nambu-Goto string world sheet action (for the gravity part) and the Kalb-Ramond string world sheet action (for the torsion part). 

Importantly though, there are entirely independent reasons for motivating this choice of the matter action \cite{Hammond, Hammond2}. We are compelled to choose a very specific classical (commutative) theory of gravitation, in order that its non-commutative version be the above gravity-torsion theory, with the action depending only on area. We have already motivated why we want the metric to be asymmetric (this is natural when the coordinates do not commute), and why we incorporate the torsion constructed from the asymmetric part of the metric (in order to include spin). One can construct the action for this classical theory minimally, and it is as usual the Einstein-Hilbert action (the volume integral over the curvature scalar), but now made from an asymmetric metric, and a metric-induced torsion tensor. The really interesting question is -- what
should be the matter source for this torsion gravity theory? Vacuum solutions of this theory have been studied and it has been shown that spherically symmetric solutions are not possible. This strongly suggests that the matter source action cannot simply be the sum  of the action terms for relativistic point masses. Rather, there should be an additional
contribution coming from the spin vector of the particles. It is then shown that such a matter source cannot be point-like, but should rather be an extended object. In fact the action for such an extended matter source can be shown to be precisely the world-sheet action for a Nambu-Goto + Kalb-Ramond string:
\begin{equation}
S_M = \mu\int \sqrt{-\gamma} \; d^2\xi + \eta \int \psi_{\mu\nu} \; d\sigma^{\mu\nu}
\label{stringaction}
\end{equation}
\begin{figure}[H]
	\centering
	\includegraphics[width=1.1\linewidth]{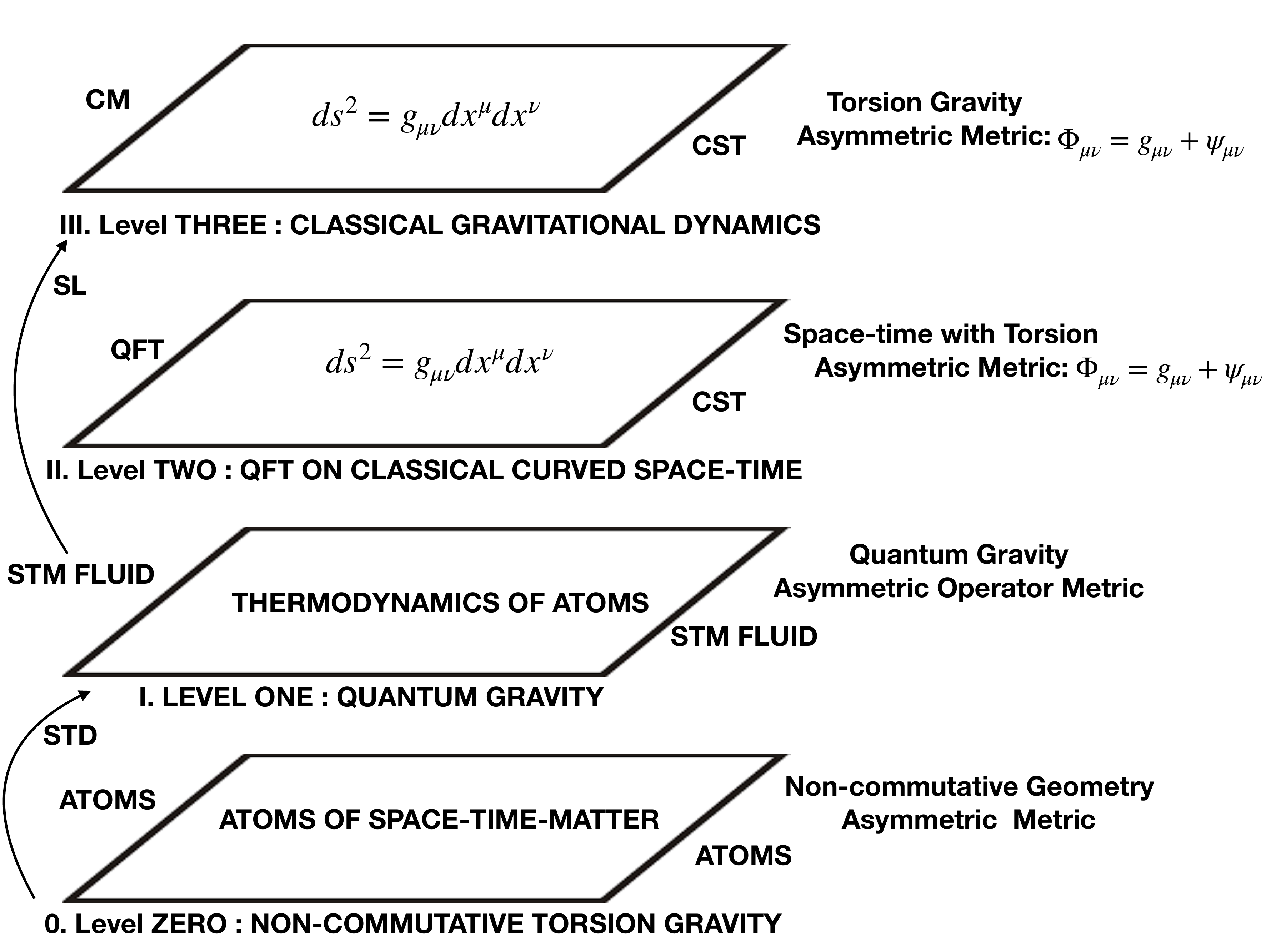}
	\caption{The four layers of gravitational dynamics. {\bf Level III} is classical gravity: a torsion gravity theory with an asymmetric metric, coupled to strings, to which classical general relativity is an 
	excellent approximation in the macroscopic realm. (CM: Classical Mechanics, CST: Classical Space-Time). {\bf Level 0}: In the microscopic realm, torsion becomes significant, but one then has to introduce non-commutativity in 
	the space-time geometry. The theory at Level 0 is non-commutative torsion gravity - a classical theory. Each space-time-matter (STM) atom is a non-commutative torsion gravity; STM atoms interact with each other through entanglement. {\bf Level I} is quantum field theory and quantum gravity. It is arrived at from Level 0 by doing the statistical thermodynamics (STD) of the STM atoms of Level 0. The transition from  Level I to Level III is via spontaneous localisation (SL). This causes the emergence of classical space-time and classical general relativity, while also explaining the quantum-to-classical transition and providing a falsifiable solution to the quantum measurement problem.  {\bf Level II}: Quantum field theory on a classical
	space-time, is a hybrid of Level I and Level III: matter is quantum but space-time geometry is classical. We explain in Section VI the conditions under which such a hybrid description is valid.}
\end{figure}
\newpage
\noindent where the two-metric is related to the four-metric in the standard way. The Kalb-Ramond
term provides a natural coupling between the string and the torsion potential. Here the asymmetric metric is $\Phi_{\mu\nu} \equiv g_{\mu\nu} + \psi_{\mu\nu}$. Thus, we have a theory of torsion gravity which couples curvature to the string world sheet action. We will refer to the theory simply as torsion gravity. The field equations are obtained by varying the action w.r.t. the symmetric and anti-symmetric parts of the metric, and the equation of motion of the string follows from the conservation of the string energy-momentum tensor. Presumably, in the limit when mass dominates over spin, this torsion gravity theory reduces to classical general relativity, as desired. We note that in the limit of classical GR (the large mass limit), the anti-symmetric part of the metric is greatly suppressed, and hence the coordinates naturally commute.

Thus, the gravitational dynamics at the classical level is torsion gravity based on an asymmetric metric, and very well approximated by general relativity in the macroscopic world. This is exhibited as Level III. in the level diagram of Fig. 1. Our fundamental theory at Level 0 is the classical but non-commutative analog of this theory, represented by the geometric action (\ref{gravityaction}) + the string action 
(\ref{stringaction}) 
\begin{equation}
S/\tilde{C}_0 = \frac{1}{L_p^2}  \int_{geom} ds^2 + \frac{1}{L^2}\int_{matter} ds^2
\label{fundaction}
\end{equation}  

The matter part of the action, having a fundamental length scale $L$, can be thought of as the material contribution to the fundamental two dimensional area. Explicitly, we assume that there is a Riemann-Cartan limit, in which we can write the total action as
\begin{equation}
S/\tilde{C}_0 = \frac{1}{L_p^2}\int_{M_{4}} R\;\sqrt{-g} d^4x +
 \frac{1}{L_1^2}\int \sqrt{-\gamma} d^2\xi + \frac{1}{L_2^2} \int \psi_{\mu\nu} \; d\sigma^{\mu\nu}
 \label{fundaction2}
\end{equation} 
assuming that there are two different length parameters $L_1$ and $L_2$ for the two
parts of the string action, each of which can be related to the string tension. Variation
of this action with respect to the symmetric and anti-symmetric metric will yield the corresponding field equations. These are non-commutative torsion gravity equations, being the non-commutative version of the torsion gravity of Level III described above.
The action has been written in dimensionless form by dividing it by a constant having the dimensions of action, denoted $\tilde{C}_0$, and to be subsequently identified with Planck's constant $\hbar$. This theory  is depicted as Level 0 in the level diagram of Fig. 1 and we simply refer to this theory as non-commutative torsion gravity. It is elegant that the emergent theory at the highest level (i.e. Level III)  and the fundamental theory at the lowest level are respectively the commutative and non-commutative analogues of each other, both being classical torsion gravity theories. In this sense, we obtain the fundamental theory at Level 0 by generalising the principle of general covariance of Level III to non-commuting coordinates. In fact, at Level 0, we can say that there is an equivalence principle, because the matter part of the action is proportional to the length (equivalently, the mass); thus the equations of motion of the string in the non-commutative space-time is independent of its length.

Remarkably, quantum field theory emerges at Level I and II as the thermodynamic 
approximation to the non-commutative classical theory at Level 0, as we explain later in this section. We thus conclude that non-commutative geometry is more fundamental than quantum theory, and in fact non-commutativity is the source for quantum behaviour. This understanding provides a geometric basis for quantum theory: quantum theory is a consequence of demanding covariance of physical laws in the presence of non-commuting reference frames. The transition from quantum theory of Level I to classical physics of Level III takes place by relativistic spontaneous localisation. Level II is a hybrid of Level I and Level III, as we will discuss later in the paper.

The fundamental action given in (\ref{fundaction}) represents an 
{\it `atom of space-time-matter'}. The string, and the space-time geometry it produces, are one and the same thing; space-time is evidently not `outside' or `inside' the string, because the string does not live in an external space-time. Each string carries with it its own space-time, so to speak, and this space-time is both non-commutative and non-local. It does not have the classical light-cone structure -- this structure only emerges at Level III, after spontaneous localisation.

In order to arrive at the string equation of motion, one needs a notion of time evolution. At Level 0, there is no classical space-time, yet one can talk of evolution with respect to the `God-given time of non-commutative geometry'. As Connes describes \cite{Connes2000}:

{\it "Noncommutative measure spaces evolve with time! 
In other words there is a `god-given' one parameter group of automorphisms of the
algebra $M$ of measurable coordinates. It is given by the group homomorphism $\delta  : R \rightarrow Out(M) = Aut(M)/Int(M) $
from the additive group $R$ to the group of automorphism classes of $M$ modulo inner
automorphisms.
I discovered this fact in 1972 when working on the Tomita-Takesaki theory  and
it convinced me that there are amazing features of noncommutative spaces which have
no counterpart in the static commutative case."} 
 
 We shall assume that time evolution at Level 0 is described with respect to this Connes time. Then the geodesic equation of motion of the string describes evolution with respect to
 the Connes time, and the equation follows from the field equations of non-commutative torsion gravity, analogously to the commutative torsion gravity case (from covariant conservation of the string energy momentum tensor). 
 
 In our theory, the universe is made up of enormously many such atoms of space-time-matter (STM). We do not know at this stage as to  what determines the total number of such STM atoms in the universe, and whether this total number is finite or infinite. Each STM atom carries its own space-time and its own matter content. The states of each STM atom live in the Hilbert space of Level 0. Consider two such atoms, labelled say 1 and 2, and consider two possible states of each atom, say $\ket{1a}, \ket{1b}$ and $\ket{2a}, \ket{2b}$. In the Hilbert space we can form the entangled state
 \begin{equation}
 \ket{\Psi_{\rm entangled}} =  \ket{1a} \ket{2a} +  \ket{1b} \ket{2b}
 \label{fundent}
 \end{equation}
 
 Entanglement describes interaction between STM atoms: interaction is entanglement, and entanglement is interaction. Clearly, the entangled state has nothing to do with classical space-time, and in fact this entanglement is more fundamental than quantum theory -- the latter emerges only at Level I and II.  Entanglement is a property of non-commutative torsion gravity, a property that is lost in the commutative limit. This entanglement involves entanglement of space-time states as well, through the STM atoms, in quite a natural way. [In this sense, entanglement also represents the `gravitational' interaction between two STM atoms].  This could have important consequences for the 
 emergence of ordinary space-time in the classical limit, for black hole entropy/evaporation, and for cosmology, as we discuss in the last section. Also, it should be apparent that since this entanglement has nothing to do with classical space-time, it will provide a resolution of the quantum non-locality puzzle and the EPR paradox. The puzzle arises not because quantum mechanics is weird or spooky, but because the classical space-time used hitherto to describe wave function collapse is an approximation to non-classical, non-local space-time, and it is the latter which is the correct arena for describing EPR correlations.
 
 It should be noted that at Level 0, the concepts of Hilbert space and entanglement have been introduced by hand. These are amongst the important assumptions of the theory at Level 0.
 
 We should also remark on the nature of configuration space for each STM atom, at Level 0. We first recall the classical torsion-gravity at Level III. Here, for the matter part, which is a closed string, the configuration variables are the string position variables. For the gravity part, in the spirit of the ADM formalism, these are the three-geometries (now built from the asymmetric metric). In going to Level 0, where non-commutativity is introduced, these same continue to be the configuration variables, except that they are now matrix-valued. Non-commutativity raises their status to that of operators / matrices. Thus the configuration variables are analogous to those in quantum general relativity, the difference now being that the commutation relations amongst them are arbitrary, and not those of quantum theory. These variables evolve in Connes time.
 
 With this, we have completed the introductory description of Level III and Level 0, which are presented in detail in Sections II and III below, respectively. We now discuss the mechanism for the emergence of Levels I and II, from Level 0, and the localisation process which enables  the emergence of Level III from Level I. 
 
 To begin with, we first recall the theory of Trace Dynamics developed by Adler and collaborators \cite{Adler:94, Adler-Millard:1996, Adler:04}.
 In this theory, space-time is flat classical  Minkowski space-time; hence there is no
 non-commutative space-time and there is no gravity. In terms of our level diagram of Fig. 1, Level 0 of trace dynamics is matrix dynamics, with each matrix describing a fundamental matter degree of freedom (which could be bosonic or fermionic). There is no Level I in trace dynamics. Level II, which is shown to emerge from Level 0, in the statistical thermodynamic limit, is quantum field theory (the equilibrium) plus fluctuations around equilibrium. Then Level III arises from Level I via spontaneous localisation. Thus trace dynamics is a theory which derives
 quantum theory from a classical matrix dynamics, and explains the quantum-to-classical transition via spontaneous localisation. The motivation is to not arrive at quantum theory by quantising a classical theory, but rather to arrive at quantum theory in a bottom-up fashion, and then to solve the quantum measurement problem by proposing that spontaneous localisation is caused by non-unitary fluctuations around the equilibrium (quantum) theory.
 
We briefly recall the principles of Trace Dynamics [TD]  - adequate details are available elsewhere \cite{Adler:04,RMP:2012,Adler:94,Adler-Millard:1996}. In brief, TD is the classical dynamics of matrices that exist on a background space-time, and whose elements are complex Grassmann numbers. The matrices could be `bosonic (B)/fermionic (F)', i.e. their elements are even-grade/odd-grade elements of the Grassmann algebra, respectively. One can construct a polynomial $P$ from these matrices, obtain its trace ${\bf P} = Tr P$, and define a so-called Trace derivative of ${\bf P}$ with respect to a matrix. Given this, a trace Lagrangian can be constructed  from the matrices (i.e. operators) $\{q_r\}$ and their time derivatives $\{\dot{q}_r\}$. Then, one can develop a classical Lagrangian and Hamiltonian dynamics for the system of matrices, in the conventional way. The configuration variables and their canonical momenta all possess arbitrary commutation relations with each other. However, as a result of the global unitary invariance of the  trace Hamiltonian, the theory possesses a remarkable conserved charge made up from the commutators and anti-commutators:
\begin{equation}
\tilde{C} = \sum_B [q_r,p_r] -\sum_F \{q_r,p_r\} 
\end{equation}

This charge, which has dimensions of action, and which we call the Adler-Millard charge, is what makes TD uniquely different from the classical mechanics of point particles. It plays a key role in the emergence of quantum theory from TD. Assuming that one wishes to observe the dynamics at a coarse-grained level, one constructs the statistical mechanics of TD in the usual way, by defining a canonical ensemble, and then determining the equilibrium state by maximising the entropy, subject to conservation laws. The canonical thermal average of the equipartitioned Adler-Millard charge takes the form (at equilibrium) 
\begin{equation}
\langle \tilde{C} \rangle _{AV}  = i_{eff}\hbar; \qquad i_{eff} = diag(i,-i,i,-i,...)
\label{canam}
\end{equation}
where $\hbar\equiv \tilde{C}_0$ is a real positive constant with dimensions of action, which is eventually identified with Planck's constant. 

Having found the equilibrium state, one can draw important conclusions from it. A general Ward identity (the analogue of the equipartition theorem in statistical mechanics) is derived as a consequence of the invariance of canonical averages under constant shifts in phase space. Its implications are closely connected with the existence of the Adler-Millard charge and its canonical average (\ref{canam}). After applying certain realistic assumptions, the canonical averages of the TD operators are shown to obey Heisenberg equations of motion, and the canonical commutation relations of quantum theory. In this sense, upon the identification of canonical averages of TD with Wightman functions in quantum field theory, one finds quantum theory emergent as a thermodynamic approximation to TD. The passage from the Heisenberg picture to the Schr\"{o}dinger picture is made in the standard manner, to arrive at the non-relativistic Schr\"{o}dinger equation.

Next, one takes account of the ever-present thermal fluctuations around equilibrium; in particular there are fluctuations in the Adler-Millard charge about its canonical value (\ref{canam}). As a result, the Schr\"{o}dinger equation picks up (linear) non-unitary stochastic correction terms. If one invokes the assumptions that evolution should be norm-preserving in spite of the stochastic corrections, and that there should be no superluminal signalling, the modified Schr\"{o}dinger equation becomes non-linear and has the generic structure of a spontaneous collapse model. One can demonstrate dynamic wave function collapse which obeys the Born probability rule. Of course the theory is not well-developed enough at this stage to uniquely pick out the CSL collapse model, or to  predict the numerical values of the CSL constants $\lambda_{CSL}$ and $r_{\text{\tiny C}}$. Moreover, the assumption of norm-preservation in the presence of stochastic corrections is ad hoc. Norm-preservation should follow from a deeper principle, yet to be discovered. Another noteworthy feature is that in order to define a thermal equilibrium state, one is compelled to pick out a special frame of reference, possibly the cosmological rest frame of the cosmic microwave background. These unresolved issues of TD will be addressed in our present paper, in Section IV and V, and in the discussion section.

Trace Dynamics (TD) is a significant example of an underlying theory from which quantum theory and collapse models are emergent. However, in TD, space-time is classical, which, as we have argued above, can only be an approximation, when matter is treated quantum mechanically. We have extended the ideas of TD to arrive at a reformulation of quantum theory without classical time, by raising time and space coordinates to the level of non-commuting matrices, as is done in TD for the canonical degrees of freedom. This program has met with some success earlier, in that it could be implemented for Minkowski space-time. The case for a curved space-time (including gravity) is developed in the present paper, and a heuristic outline is available in earlier work \cite{Singh:2012, Singh:2019a}. Below, we briefly recall our earlier results for the generalised  TD on a non-commutative Minkowski space-time  \cite{Lochan-Singh:2011, Lochan:2012}.

On a non-commutative Minkowski space-time, with $(\hat{t},\hat{x},\hat{y},\hat{z})$ as non-commuting operators having arbitrary commutation relations, we define a trace proper time as follows 
\begin{equation}
ds^2 = Tr d\hat{s}^2 \equiv Tr [ d\hat{t}^2 - d\hat{x}^2 - d\hat{y}^2 - d \hat{z}^2 ] 
\label{nsr}
\end{equation}

This line-element can be shown to be invariant under Lorentz transformations. Matter degrees of freedom `live' on this non-commutative space-time, and one can define a Poincar\'e invariant dynamics, by first introducing a four-vector $\hat{x}^{\mu} = (\hat{t},{\bf \hat{x}})$ and defining a four velocity as $\hat{u}^{\mu}=d\hat{x}^{\mu}/ds$. Lagrangian and Hamiltonian dynamics can then be constructed in the spirit of Trace Dynamics, using the trace proper time to define evolution. As before, there exists in the theory, as a consequence of global unitary invariance, a conserved Adler-Millard charge for the matter degrees of freedom $\hat{y}^\mu$:
\begin{equation}
\hat{Q} = \sum_{r\in B} [\hat{y}_r,\hat{p}_r] - \sum_{r\in F} \{\hat{y}_r,\hat{p}_r\}
\end{equation}

The important generalization is that there is associated, with every degree of freedom, apart from the canonical pair $(\hat{q}, \hat{p})$, a conjugate pair $(\hat{E},\hat{t})$,  where $\hat{E}$ is the energy operator conjugate to $\hat{t}$. The metric above, though Lorentz invariant, does not admit a light-cone structure, nor the point structure of ordinary space-time. This is what allows the recovery of a quantum theory without classical time. 

From this point on, the construction parallels that in TD, namely an equilibrium statistical thermodynamics of the underlying classical theory is constructed, with coordinate time $\hat{t}$ now an operator, and evolution being described with  respect to trace proper time ${s}$. The quantum commutators emerge at the thermodynamic level, with the added feature that there is now an energy-time commutator as well. In the non-relativistic limit one obtains the generalised Schr\"{o}dinger equation:
\begin{equation}
i\hbar \frac{d\Psi}{ds} = H\Psi(s)
\label{gqd}
\end{equation}
where the configuration variables now include also the time operator $\hat{t}$. It is important to emphasize that the configuration variables now commute with each other. We call this a Generalised Quantum Dynamics (GQD), in which there is no classical space-time background, and all matter and space-time degrees of freedom have operator status. This is the sought for reformulation of quantum theory which does not refer to classical time \cite{Lochan:2012}, i.e. Level I.

This is the construction which takes us from Level 0 to Level I, when there is no gravity included, and when there is a global operator Minkowski space-time at Level 0. The transition from Level I to Level III takes place via our recently proposed mechanism of relativistic spontaneous localisation. \cite{Singh:2018, Singh:2019}. We showed in this work that classical space-time and classical dynamics emerge from Level I through spontaneous collapse of the wave function.  We also explained therein how the hybrid Level II (quantum theory on classical Minkowski space-time) is arrived at.

Thus, in previous work, we have shown how to develop a quantum field theory (on Minkowski space-time) which does not refer to classical time, by using the techniques of trace dynamics and relativistic spontaneous localisation. This gives the four levels of dynamics as in Fig. 1 except that gravity is not included. In the present paper, we carry out the task of including gravity in these four levels. and this presents us with a falsifiable candidate theory of quantum gravity, i.e. non-commutative torsion gravity.  Section IV describes how a quantum gravity theory emerges at Level I from the statistical thermodynamics of the STM atoms of Level 0. Section V describes how spontaneous localisation takes us from Level I to the classical gravity world of Level III, and Section VI explains how the hybrid Level II is obtained. Section II below describes torsion gravity, and Section III describes its non-commutative version.
 
\section{Level III: Torsion gravity based on an asymmetric metric}
This discussion is largely based on the review article by Hammond \cite{Hammond}. 
Given a Lorentzian space-time four manifold, we denote the asymmetric metric as
\begin{equation}\Phi_{\mu\nu} \equiv g_{\mu\nu} + \psi_{\mu\nu}\end{equation}
and the asymmetric connection as
\begin{equation}
\tilde{\Gamma} \indices{_{\mu \nu}^{\sigma}} = \Gamma \indices{_{\mu \nu}^{\sigma}} + S \indices{_{\mu \nu}^{\sigma}} \
\label{3a}
\end{equation}
The symmetric part of the connection (Christoffel symbols)  is related to the symmetric part of the metric as usual, and we define the anti-symmetric part of the connection (i.e. the torsion tensor, which in this case is completely  anti-symmetric) from the  anti-symmetric part of the metric:
\begin{align}
S \indices{_{\mu \nu \sigma}} = \psi \indices{_{[\mu \nu, \sigma]}} = \frac{1}{3} (\psi \indices{_{\mu \nu, \sigma}} + \psi \indices{_{\sigma \mu, \nu}} + \psi \indices{_{\nu \sigma, \mu}})
\label{2a}
\end{align}
The gravity-torsion part of the action is the Einstein-Hilbert action for an asymmetric metric and asymmetric  connection
\begin{align}
\mathcal{S}_g = \frac{1}{2 \kappa} \int d^4x \sqrt{-\Phi} \, R(\tilde{\Gamma}) \,,
\label{2b}
\end{align}
Varying this gravity-torsion part of the action with respect to the symmetric and anti-symmetric part of the metric gives the following vacuum field equations.
\begin{equation}
G^{\mu \nu} - S \indices{^{\mu \nu \sigma}_{; \sigma}} - 2 S \indices{^{\mu}_{\alpha \beta}} S^{\nu \alpha \beta} = 0 \,,
\label{4a}
\end{equation}
\begin{equation}
S \indices{^{\mu \nu \sigma}_{; \sigma}} = 0 
\label{4b}
\end{equation}
where the semi-colon derivatives are Christoffel covariant derivatives.

It can be shown that these vacuum equations do not admit spherically symmetric solutions. This is a hint that a point mass by itself cannot be the source for such a torsion-gravity field. A vectorial matter source, denoted $\xi^\mu$, has to be invoked as well, and spin is the most likely candidate for such a source. One can write the action for a relativistic mass-spin object as an extended source, as follows:
\begin{align}
\mathcal{S}_M = \sum_{n} \bigg[ m_n c^2 \int d \tau_n \, \sqrt{\frac{dx^{\mu}_{n}}{d\tau_n}\frac{dx^{\nu}_{n}}{d\tau_n} g_{\mu \nu}} + \frac{c^2}{2} \int d\tau_n \, \xi_{n}^{\mu} \frac{dx^{\nu}_{n}}{d\tau_n} (2 \psi_{\mu \nu} + g_{\mu \nu}) \bigg] 
\label{6a}
\end{align}
where the sum over $n$ represents the sum over all of the infinitesimal regions of the particle. The fact that there must be an inherent structure to the source is extensively motivated in \cite{Hammond}. The matter action extensively relies on the introduction of the intrinsic vector quantity $\xi^{\mu}$ which is identified with the spin vector and is the source of spacetime torsion. 

The variation of the total  action $(S_g + S_M)$ with respect to the asymmetric metric $\Phi_{\mu \nu}$ yields the asymmetric field equations:
\begin{align}
G^{\mu \nu} - 3S \indices{^{\mu \nu \sigma}_{; \sigma}} - 2 S \indices{^{\mu}_{\alpha \beta}} S^{\nu \alpha \beta} = \kappa T^{\mu \nu} \,.
\label{7a}
\end{align}

The symmetric part of (\ref{7a}) are essentially the gravitational field equations:
\begin{align}
G^{(\mu \nu)} - 2 S \indices{^{\mu}_{\alpha \beta}} S^{\nu \alpha \beta} = \kappa T^{(\mu \nu)} \,,
\label{8a}
\end{align}
while its antisymmetric part are the torsional field equations:
\begin{align}
S \indices{^{\mu \nu \sigma}_{; \sigma}} = - \kappa J^{\mu \nu} \,,
\label{9a}
\end{align}
where $J^{\mu \nu} \equiv \frac{1}{2} T^{[\mu \nu]}$.

The energy-momentum tensor is determined by requiring that it be compatible with the above matter action; this is discussed in detail in \cite{Hammond}. 

The most interesting aspect of the theory with the ansatz (\ref{2a}) is that it brings to light a connection between the antisymmetric field of string theory and spacetime torsion. This connection is not apparent from a Poincar\'e gauge gravity perspective, as is extensively discussed in \cite{Hammond}. In context of a metric theory of gravity ($ g_{\mu \nu;\alpha} = 0$), this idea was well refined in \cite{Hammond2} which predicts the precise form of the low-energy effective Lagrangian of string theory. In addition, the postulated matter action (\ref{6a}) based on the intrinsic spin vector $\xi^{\mu}$ assumed a structure for its source by virtue of a summation over infinitesimal parts of the body. Furthermore, some general conditions like $\sum_{n} \xi^{\mu}_{n} = 0$ are imposed as well. Such conditions are naturally seen to arise when the source is taken to be a closed string, as elucidated extensively in \cite{Hammond}. Due to these common features, it is a well motivated decision to adopt a material action for a string. Thus the matter action is now given by:
\begin{align}
\mathcal{S}_{M} = \mu \int \sqrt{- \gamma}\; d^2 \chi + \eta \int \psi_{\mu \nu}\; d \sigma^{\mu \nu} \,,
\label{10a}
\end{align}
with $\gamma$ being the determinant of the two dimensional metric. The first term is the Nambu-Goto term and gives rise to a symmetric energy-momentum tensor. The second term is the Kalb-Ramond term and introduces torsion into the theory. It provides a very natural coupling between the string and the torsion potential. 

The full action is given by
\begin{align}
\mathcal{S} =  \frac{1}{2 \kappa} \int d^4x \sqrt{-\Phi} \, R(\tilde{\Gamma}) + \mu \int \sqrt{- \gamma}\; d^2 \chi + \eta \int \psi_{\mu \nu}\; d \sigma^{\mu \nu} \,,
\label{fullacn}
\end{align}

The variation with respect to the metric yields the same field equations as above, but with the energy-momentum tensor now determined from the string action, as follows:
\begin{equation}
G^{\mu \nu} - 3S \indices{^{\mu \nu \sigma}_{; \sigma}} - 2 S \indices{^{\mu}_{\alpha \beta}} S^{\nu \alpha \beta} = \kappa T^{\mu \nu} \,.
\label{7a}
\end{equation}
\begin{equation}
G^{(\mu \nu)} - 2 S \indices{^{\mu}_{\alpha \beta}} S^{\nu \alpha \beta} = \kappa T^{(\mu \nu)} \,,
\label{8a}
\end{equation}
\begin{equation}
S \indices{^{\mu \nu \sigma}_{; \sigma}} = - \kappa \frac{1}{2} T^{[\mu \nu]} \,,
\label{9a}
\end{equation}
\begin{equation}
T^{\mu\nu} = \frac{\mu}{\sqrt{-\Phi}} \int d^2 \xi \; \sqrt{-\gamma}\; \delta (x- x(\xi))  \; x^\sigma_{,a}\; x^{\nu}_{,b} (\gamma^{ab} + \lambda\epsilon^{ab})
\end{equation}
with $\lambda = 2\eta / \mu$. These are field equations of the theory which we call torsion gravity. The equation of motion of the string is discussed in detail in \cite{Hammond}.

One can in principle set up the ADM formalism for this torsion gravity theory, and identify the canonical metric variables and their conjugate momenta, as well as the Hamiltonian and the constraint equations. We leave this investigation for future work.

We have described above the basic elements of the torsion-gravity-string theory that lives on Level III. In the macroscopic limit, it is reasonable to assume that mass dominates spin. Hence the antisymmetric metric and the torsion can be neglected, and the theory reduces to general relativity. The string action, being equivalent to the matter action above, can be written as that for a relativistic point mass, since spin is being neglected.

In the next section we build the non-commutative generalisation of this theory, so as to arrive at the STM atom of Level 0. Thus we do not quantize the above classical theory; rather we 
`non-commutativise' it. While the classical theory above describes the space-time of the whole universe (after summing over all matter degrees of freedom), the non-commutative version of the above theory describes an individual space-time-matter atom. As if each atom were its own universe. When one develops the statistical mechanics of a large number of such atoms, then quantum theory emerges at thermodynamic equilibrium.  That is, the `quantised' version of the above classical theory emerges as the coarse-grained approximation of the non-commutativised theory. Classical space-time geometry and the commutative torsion gravity presented above emerges at Level III from this quantum theory after spontaneous localisation, and after averaging over the space-times of individual STM atoms. It is also worth asking if there is a direct mechanism to go from the non-commutative theory of Level 0 to its commutative version at Level III, without having to go through Levels I and II.

\section{Level 0: Non-commutative torsion-gravity: atoms of space-time-matter}

\subsection{An overview of non-commutative geometry}

\indent Perhaps one of the most fertile branches of mathematics that has emerged in recent years is that of non-commutative geometry. Although inspired by his work with Dixmier on C* algebras and the work of Atiyah-Singer on topological K-theory, non-commutative geometry is largely due to the original work of Alain Connes, its basic structure and scope laid out in Connes's comprehensive book \cite{connes1994noncommutative} and review article \cite{Connes2000}.

Connes's celebrated Reconstruction Theorem -- the recovery of an underlying manifold from an algebra of functions on it -- builds upon the work of Gel'fand and Naimark, who showed that there is a complete isometry between the geometry of a topological space (metric space, phase space, etc.) and the (commutative) C* algebra of continuous functions $f:M\rightarrow \mathbb{C}$. In other words, there is in a sense an equivalence, or rather \textit{duality}, between the geometrical and algebraic descriptions \cite{petitot_2009}. This is of course useful in the case of quantum mechanics, where the spectra of observables is given by the operator algebra rather than a set of points on a classical spacetime. These operator algebras (Hilbert space operators, von Neumann algebras) are, importantly, non-commutative. Connes' non-commutative geometry (NCG) represents a greatly successful programme of geometrising the algebra of non-commutative operators, in effect asking the question: ``In the spirit of Gel'fand, can we recover the unique and completely isometric geometrical space defined by the non-commutative C* algebra of bounded operators acting on a Hilbert space?"

Connes invents various algebraic-geometric tools to describe this new non-commutative ``space" under consideration, including an abstract non-commutative (pseudodifferential) calculus, a method of Dixmier traces to solve integrals, and crucially, the notion of a \textit{spectral triple}. A spectral triple $(\mathcal{A}, \mathcal{H}, \mathcal{D})$ is a Fredholm module that consists of a unital, involutive algebra $\mathcal{A}$, a Hilbert space $\mathcal{H}$, and a linear Hermitian operator $\mathcal{D}$ acting on $\mathcal{H}$. Connes' Reconstruction Theorem demonstrates that out of a commutative spectral triple (a spectral triple with a commutative algebra $\mathcal{A}$), one can recover the underlying manifold, with associated vector bundles and connections. Similarly, in the almost commutative case (used in constructing the non-commutative Standard Model) -- the tensor product of a commutative spectral triple and a general finite one -- there  exists a generalisation of this theorem. 

With a seminal contribution to the physics and philosophy of the notion of ``distance", Connes (and Chamseddine) were able to demonstrate the Spectral Action principle -- that the spectrum of $D$ -- suitably defined -- uniquely fixes the physical action. With the explicit definition of $ds = D^{-1}$ (where $D$ is the Dirac operator) and the universally true commutation relation $[[D,f],g] = 0$ (for all $f,g \in \mathcal{A}$) Connes laid down the first two axioms of his geometry (thus far, we have only satisfied the requirements for a \textit{commutative} geometry); other axioms include mathematical considerations such as smoothness and finiteness \cite{Connes:1996gi}. For the full departure into NCG, the second of these is modified with $g^0$ replacing $g$ where $g^0$ is the canonical antilinear isometric involution $Jg^*J^{-1}=g^0$ from Tomita-Takesaki theory (note, $f$ and $g^0$ commute). 

The Tomita-Takesaki theorem is a powerful and relevant result that does not exist in the commutative analogue. Indeed, it requires  that the geometry  be non-commutative. It asserts that there is a (as Connes puts it, ``god-given") one-parameter group of inner automorphisms of the algebra $\mathcal{A}$ which gives us a universal parameter according to which non-commutative spaces evolve -- which Connes identifies with evolution time. As we will see, this becomes important when we seek to work with
 matrix - or operator - time (such as in Trace Dynamics or Horwitz-Stuckelberg theory), where the evolution equations require a universal (and otherwise unmotivated) $\tau$ parameter.

The crucial physical insight offered by Connes is the new measure of distance -- in essence, this is where the non-commutativity truly comes in. The identification of the distance measure $ds$ as the inverse of the Dirac operator $D$ with $D = \gamma^\mu \bigtriangledown_\mu$ represents a novel -- and \textit{physical} -- measure of distance on a geometry (and well-motivated by physical intuition -- this is nothing but the Feynman propagator) and creates tension between $ds = D^{-1}$ and the spatial coordinates $a \in A$. To quote Connes \cite{Connes2000} --``it is precisely this lack of commutativity between the line element and the coordinates on a space that will provide the measure of distance". With Connes's redefinition of the differential (motivated by the quantum evolution equation), we now have $df = [D, f]$ giving the differential for any $f \in A$ in the NCG scheme.

In this scheme, the geodesic distance between two points $(x,y)$  on a manifold is formally rewritten as \cite{petitot_2009}: 

\begin{equation*}
d(x,y) = \text{Inf} L(\gamma) = \text{Inf} \int_x^y ds = \text{Inf} \int_x^y \sqrt{g_{\mu\nu} dx^\mu dx^\nu}
\end{equation*}
where the $L(\gamma)$ are the lengths of the paths $\gamma:x\rightarrow y$. From this, it can then be shown (Connes' `distance formula') that the dual algebraic version of this definition is in terms of the supremum on $x \in M$ of the norms on the tangent spaces $T_xM$: 

\begin{equation*}     
d(x,y) = \text{Sup} \{|f(y) - f(x)|; f \in A, ||\bigtriangledown f||_\infty \leq 1\}
\end{equation*}

The $\bigtriangledown$'s, as we have seen, are simply given by the commutators, $[D, f]$, and hence this definition helps us define the metric through $D$.

A note to be made here -- due to the construction of the abstract geometry from the operator algebra on the basis of the spectral triple, $\{\mathcal{A},\mathcal{H},\mathcal{D}\}$, we no longer have the Riemannian manifold of space-time as a fundamental entity, but rather, a derived concept, constructed out of the $D^{-1}$ which smoothly cover ordinary space. This means that we no longer consider the 
configuration - or coordinate - space description as fundamental; indeed the spatial points $\{\bold{x}\}$ no longer have any meaning outside of the spectra of the corresponding operator in our algebra (such as $\hat{x}$). In other words, the $\{\bold{x}\}$ are merely a collection of eigenvalues, and are no longer to be used in the classical sense to define intervals $\Delta \bold{x} = \bold{x_2} - \bold{x_1}$. This means, among other things, that nowhere in the NCG scheme do we refer to the commutativity of the coordinates themselves, they could indeed commute or not commute depending on the specific theory at hand. Every spatial interval $ds$ is given by the commutator with the Dirac operator, $[D, x]$, while the relativistic line element is simply in terms of these commutators and the metric (at this stage no more than a matrix of numbers): $ds^2=[D,x^\mu]*g_{\mu\nu}[D,x^\nu]$. 

In effect, the only measure of distance here is the commutator with the Dirac operator. With this measure of distance, we have the important result in NCG \cite{Connes:1996gi}: If $A = C^{\infty}M$ where $M$ is a smooth, compact manifold and all the NCG axioms are satisfied, then there exists a unique Riemannian metric $g$ on $M$ defined as above, and 
\begin{equation*}
\intbar ds^2 = \frac{-1}{48\pi^2}\int_{M4}R\sqrt{g}d^4x 
\end{equation*}

Connes motivates this result on symmetry grounds, while Kalau \& Walze, and separately, Kastler, demonstrates this explicitly, using the Wodzicki residue. The crucial point here is tha the spectral action can be formulated as an expansion. As mentioned, the spectral action is a functional on the space of spectral triples. More explicitly, it is a regularised heat kernel expansion of $D$. As the Einstein-Hilbert action is a functional on the space of Riemannian manifolds, the spectral action is a functional on a general non-commutative space. For ordinary Riemannian spaces, the spectral action reduces to the EH action, plus integrals over higher curvature invariants (interestingly, Chamseddine claims that a generalised 2-spectral triple can be used, analogously, to describe the propagation of a superstring on a generalised Riemannian space with a twisted string structure. The analogue of $D$ here is the Dirac-Ramond operator). 

Two different mathematical tools were unified by Connes, in the attempt to calculate integrals of members of the operator algebra, $T \in A$, the Dixmier trace and the Wodzicki residue. The integral $\intbar T$ is defined as the integral of a first-order infinitesimal in $T$, and is identified with the coefficient of logarithmic divergence in the trace of $T$. This is due to the standard analysis by Dixmier \cite{Connes2000}, which allows us to compute singular traces on a space of linear operators. This can be applied to the \textit{dimension spectrum} of a non-commutative geometry, defined as the subset of the complex plane $\mathcal{C}$ where the spectral functions have singularities. Assuming the spectral functions have at most simple poles, the residues at the poles are given by the Dixmier traces. This is simply an extension of the Wodzicki residue of pseudodifferential operators on manifolds. In other words,
\begin{equation*}
\intbar T = \text{Res}_{s=0} Tr(T|D|^{-s})
\end{equation*}
giving us the general method of calculating integrals in the NCG framework \cite{connes_2006}.

Kalau and Walze \cite{kalau_walze_1995} successfully use the method of calculating Wodzicki residues to derive the Einsein-Hilbert action. From the observation that the metric structure is fully encoded in the Dirac operator of a K-cycle, they set out to define the curvature tensors and the EH action from the logarithmic divergent part of $\text{tr}(D^{-n+2})$. They find that the Wodzicki residue of $D^{-n+2}$ picks out the second heat kernel coefficient and results in the EH action. Kastler \cite{kastler_1995} derives independently the same result. Both parties start with the Lichnerowicz formula, which expresses the Dirac operator in terms of the Clifford connection, the metric, and the scalar curvature \cite{Ackermann:1994pg}: 
\begin{equation}
D^2 = -g_{\mu\nu}( \tilde{\bigtriangledown}_\mu \tilde{\bigtriangledown}_\nu - \Gamma^\alpha_{\mu\nu} \tilde{\bigtriangledown}_\alpha) + \frac{1}{4}R
\end{equation}

After some calculation (see \cite{kastler_1995} and \cite{kalau_walze_1995}), this yields the desired result $\intbar ds^2 = S_{GR} = \frac{1}{12}\int d^n x \sqrt{g}R$. The details of the derivation are somewhat tedious, but it suffices to note that the first group of terms end up being proportional to the scalar curvature as well, allowing us to end up with an overall multiple of the curvature.

But what of torsion? The preliminary studies into torsion in the NCG framework were done by Chamseddine et al. in \cite{chamseddine_felder_frolich_1993,Chamseddine:1995}, where they first formulate the expression for torsion in NCG but then go on to use the no-torsion condition to derive the classical EH action. Others, such as in \cite{Ciric:2016isg}, have looked at the Poincar\'e gauge gravity formalism in the NCG context, and the reader is also referred to \cite{beggs_majid_2011} for an excellent review of compatible connections.

Kalau and Walze \cite{kalau_walze_1995} do go on to show that any torsion term added to the Clifford connection of the form $\bigtriangledown^\epsilon_\mu + T_\mu$ eventually results in the corresponding tensor added to the action, i.e. a resulting action of the form $S_{GR} = \int d^n x \sqrt{g}(\frac{1}{12}R - \frac{3}{4}t_{abc}t^{abc})$. However, the consideration of Hammond-like torsion has not yet been done, in the NCG framework.

From the above summary, we identify two possible points of departure from standard NCG theory. Firstly, we note that, in a generalisation of Kalau's calculation for a connection with torsion, we can also have a term coming in from the antisymmetric part of the metric itself, as well as the tensor from which the connection due to torsion is derived. For the explicit form of this term, the calculations of Kalau and Kastler will have to be carried out explicitly, with a corresponding torsion tensor of the Hammond type. This is a program for future work.

Finally, as explained, the only measure of distance in NCG comes from the commutator of the $\{x^\mu\}$ with the Dirac operator. This allows for the possibility that the $\{x^\mu\}$ themselves can be taken to be non-commuting variables, as in the theories of Horwitz-Stuckelberg \cite{Horwitz:2006} or generalised trace dynamics \cite{Lochan-Singh:2011}.

\subsection{Non-commutative torsion gravity as a matrix dynamics}
We propose that gravity is classical (but non-commutative) at the Planck scale, and that the equivalence principle continues to hold at Level 0, in the sense described in the previous section.
Non-commutativity makes a profound difference to ordinary commutative classical dynamics and gravitational dynamics, such as the one described in the previous 
section.
The non-commutative theory is in a sense more quantum than quantum theory, with the latter emerging only after coarse-graining from the former. `More quantum' in the sense that the commutation relations between dynamical variables are completely arbitrary at Level 0.

While it remains to be shown, we will assume for now that the non-commutative analog of the above torsion-gravity theory (\ref{fullacn}) can be described by an area action of the Connes kind, as given in (\ref{fundaction}). We note that the gravitational constant $G$ has been replaced by the Planck length $L_{pl}$, from which $G$ will emerge subsequently, after the Planck constant $\hbar$ emerges. Evolution is described w.r.t. the Connes time $\tau$. Space-time coordinates do not commute, and obey arbitrary commutation relations.

Further, we will assume, subject to confirmation by further investigation, that this
non-commutative gravity can equivalently be written as a matrix dynamics, in the sense of trace dynamics. The canonical configuration and momentum variables for the torsion gravity theory now become operator/matrix valued, and obey arbitrary commutation relations with each other. In particular, the components of the asymmetric metric are now operator valued. The Lagrangian density of the theory is now the trace of the operator polynomial corresponding to the Lagrangian in (\ref{fullacn}), as in trace dynamics. Essentially, the classical matrix dynamics corresponding to torsion gravity is constructed exactly as in trace dynamics; it is just that we have now specified the Lagrangian. Differentiation with respect to operator valued canonical variables is defined using the concept of trace derivative. An invariant volume element, as required for constructing the action in (\ref{fullacn}), is made from the determinant of the trace of the matrix valued metric -- this is the proposal made by Adler in \cite{Adler2014}, which we implement here. 

This motivates us to write the matrix dynamics version of the action (\ref{fullacn}) as
\begin{align}
\mathcal{S/{\tilde{C}}} =  \frac{1}{L_{pl}^2} \int d^4x \; \sqrt{-Det(Tr \Phi)} \, Tr[{\bf R(\tilde{\Gamma})}] + \frac{1}{L_1^2} \int \sqrt{- Det{(Tr\gamma)}}\; d^2 \chi + \frac{1}{L_2^2} \int Tr[{\bf \psi_{\mu \nu}\; d \sigma^{\mu \nu}}] \,,
\label{fullacn2}
\end{align}
This is the action which we assume to be equivalent to the Connes action (\ref{fundaction}), though this remains to be shown.

This action describes a single space-time-matter atom; each atom is a universe in itself. Time evolution is described w.r.t. Connes time, and the non-commutative field equations are obtained by varying this action w.r.t. the metric. These are the Lagrange (equivalently, 
Hamilton) field equations for a space-time-matter atom. These are operator equations, being the operator analogue of the field equations (\ref{7a}). The eigenstates of these operator equations live in the Hilbert space, where there is of course no classical 
space-time as yet.

We assume the full Hilbert space to be populated by a large number of such STM atoms, with the full action being the piecewise sum of the individual actions for each STM atom, and possessing a conserved charge -- the Adler-Milard charge $\tilde{C}$ -- as a result of global unitary invariance of the total Lagrangian. Using the states of individual STM atoms, one can form entangled states, with the entanglement describing interaction.

\section{Level I: Statistical thermodynamics for non-commutative torsion gravity: emergence of quantum gravity}
Given the Hamiltonian for each STM atom at Level 0, one develops the statistical thermodynamics of a large number of such atoms, as is done in the theory of trace dynamics. The total Hamiltonian is the sum of Hamiltonians of individual STM atoms, and 
at thermodynamical equilibrium, the conserved Adler-Millard charge is equipartitioned over all the degrees of freedom, with the constant equipartitioned value per each degree of freedom assumed to be equal to the Planck constant $\hbar$. Thus, in $\hbar$ we have the third fundamental constant of the theory, after Planck length and speed of light, and then Newton's gravitational constant $G$ is defined as $G \equiv L_{pl}^2 \; c^3 / \hbar$. As in trace dynamics, the canonical thermal averages of the dynamical variables obey the standard commutation relations of quantum field theory, and they also obey the Heisenberg equations of motion. Evolution is given by  Connes time, and there is still no classical space-time. The thermal bath in which the STM atoms are assumed to be in equilibrium requires a universal time evolution variable to be specified, and we assume that to be Connes time. Since there is no classical space-time at this level, we do not see this as any violation of Lorentz invariance. By the time Lorentz invariance emerges at Level III, non-commutativity is lost, and hence the Connes time is lost too.

At the thermodynamic equilibrium in the emergent level, in the functional Schr\"{o}dinger picture, one gets a Wheeler-DeWitt like equation, for the state $\Psi_i$ of the gravitational (i.e. asymmetric metric) and matter degrees of freedom of the $i$-{\it th} STM atom:
\begin{equation} 
i\hbar \frac{\delta \Psi_i}{ \delta \tau} = H_i \; \Psi_i 
\end{equation}
where evolution is with respect to the Connes time $\tau$. $H_i$ is the canonical trace Hamiltonian constructed from the action (\ref{fullacn2}).  Because there is no interaction (except entanglement) between the STM atoms, their individual Hamiltonians decouple from each other.
It will be very interesting to investigate the possible connection of the quantum gravity at this level, with loop quantum gravity.
In the emergent commutative limit at Level III, this non-commutative measure of evolution is lost, space-time emerges, and Lorentz invariance is realised. 

One is still in the same Hilbert space as that of Level 0, but we now have a coarse-grained view of this space. We are seeing the approximate equations satisfied by the STM atoms, in thermodynamic equilibrium, and these are the equations of quantum gravity. Quantum gravity is an emergent phenomenon, arising from an underlying non-commutative classical dynamics operating in the Hilbert space at the Planck scale. As before, one can make entangled states from solutions of the Wheeler-deWitt like equations for different STM atoms.

Given the thermodynamic origin of quantum general relativity, it should be possible to express the theory in terms of conventional thermodynamic variables, exploiting their conventional relation to the Boltzmann entropy and the partition function. This also helps understand the thermodynamic nature of Einstein equations \cite{Jacobson1995, paddy} and of black holes.

As in trace dynamics, statistical fluctuations about equilibrium play an extremely significant role in this theory. These fluctuations prevent the equipartitioning of the 
Adler-Millard charge, and result in departures from quantum theory and quantum general relativity.  The impact of fluctuations on the averaging of $\tilde{C}$ can be represented as stochastic corrections to its ensemble average. This in turn results in stochastic modification of the emergent Heisenberg equations of motion. Equivalently, the
 Wheeler-deWitt like functional Schr\"{o}dinger equation gets endowed with stochastic modifications, because the fluctuations come from corrections to the Adler-Millard charge.
 
 While the STM Hamiltonians for different atoms decouple from each other, the fluctuations couple different STM atoms, because the corrections to $\tilde{C}$ depend on all the atoms, and not necessarily in a separable way. Thus the stochastic 
 Wheeler-deWitt like equation takes the form
 \begin{equation}
 i\hbar \frac{\delta \Psi (q_1,q_2,...)} {\delta \tau}  = \left(\left[ \sum_{i} H_i(q_i)\right] + {\mathcal K} (q_1,q_2,...) \right) \Psi(q_1,q_2,...)
 \end{equation}
 where ${\mathcal K}$ represents stochastic corrections to the quantum gravity equation.

 These modifications in principle cause stochastic wave vector reduction, as in collapse models and trace dynamics, except that we now have the advantage that the Lagrangian is known explicitly. It should hence be possible to determine the values of the collapse model parameters from first principles. This investigation is under progress. For the present, we will restrict ourselves to a GRW type relativistic spontaneous localisation model, and explain in the next section how it takes us from Level I to the classical Level III. This quantum-to-classical transition is responsible for the emergence of classical space-time and classical gravity.

 The above equation has the remarkable property that when the fluctuations are negligible, we have quantum theory, and  the STM atoms are free and do not interact with each other gravitationally. Although every atom has its own operator space-time and curvature-torsion, their mutual interaction is only through entanglement. On the other hand, when the fluctuations are completely dominant over the Hamiltonian evolution, spontaneous localisation takes place, resulting in the emergence of space-time and of gravity, and entanglement is lost. Entanglement and gravity are complimentary -- the former becoming more important for small masses, and the latter for large masses. However, together, they yield the above equation which as we have argued earlier, obeys an equivalence principle, and is hence in that sense mass independent. In the classical limit this equivalence principle is preserved, and we clearly see that by itself quantum theory violates the equivalence principle.

\section{Level III: Spontaneous Localisation: from quantum gravity to classical general relativity}
Spontaneous localisation has been proposed in the literature as a mechanism to explain the quantum-to-classical transition in quantum physics, and to provide a falsifiable solution to the quantum measurement problem \cite{Ghirardi:86, Ghirardi2:90}. However, we have shown that the idea of spontaneous collapse has a much greater relevance than the measurement problem. A relativistic generalisation of spontaneous localisation suggests that space-time in itself arises from this collapse process, and the solution of the measurement problem is simply a special case of this phenomenon \cite{Singh:2018, Singh:2019}. 
What was not considered in our earlier work was the origin of gravity. Below we outline how gravity emerges at Level III, starting from Level I. The mathematical details remain to be worked out, but the overall picture is the same as in the earlier work
\cite{Singh:2018, Singh:2019}. 

Consider a state $\Psi$ of $N$ STM atoms in the Hilbert space at Level I, labeled by the space-time operator coordinates $\hat{x}^{i}_n$ of the various atoms. Thus 
$\Psi= \Psi (\tau, {\hat x}^{i}_1, \hat{x}^{i}_2, ...., \hat{x}^{i}_N)$. 
At random $\tau$ time, the matter part of the $n$-th atom undergoes spontaneous localisation to some random eigenvalue $x^{i}_n$ of its operator coordinate. This localisation process is defined by the jump operator. The (non-matter) curvature-torsion part of the STM atom stays uncollapsed. Macroscopic states are those which involve entanglement of many atoms - these undergo very  rapid spontaneous collapse, and are responsible for the emergence of classical space-time.  The classical field equations of Level III arise as follows. Upon spontaneous collapse, each atom obeys the field equations of commutative torsion gravity as given above in Section II. It appears that the average space-time produced by many STM atoms is somehow the average of the individual space-times of the various STM atoms: the metric, connection, and curvature are all averaged, e.g. the curvature scalar $R$ of the emergent classical space-time is the average of the curvature scalars $R_n$ of the individual atoms:
$R = <\{R_n\}>$. With this assumption, the field equation of the emergent universe is the same as the equation (\ref{7a}) of classical torsion gravity, with the matter source being given by energy-momentum coming from Eqn. (\ref{6a}), the sum over $n$ now representing the sum over many particles. When spin can be neglected, these reduce to the field equations of classical general relativity. We bypass the need to know explicit expressions for the string lengths $L_1, L_2$ in so far as getting the emergent theory is concerned.

Uncollapsed objects stay on Level I -- this means that fluctuations are not significant for them. For such one or more STM atom(s), dynamics is described by the Wheeler-deWitt like equation (without fluctuations). There is no background space-time, nor a gravitational interaction, but entanglement is possible.

\section{Level Ii: A hybrid of Levels I and III: Quantum field theory on a classical curved background}
Level II describes quantum field theory on a classical curved background, and in particular on a Minkowski background. In our earlier work we have discussed how this hybrid level arises from making suitable approximations at Level I, when the background is Minkowski. A similar reasoning holds now as well. Strictly speaking, the quantum fields of Level II should be described at Level I, because as we have argued earlier, one cannot fundamentally have matter as quantum and simultaneously space-time as classical. Except by making appropriate approximations and assumptions at Level I. These approximations are as follows: a) The curvature produced by the STM atoms at Level I is neglected. b) In the resulting operator space-time, the anti-symmetric part of the metric is suppressed, so that one is left only with an operator  Minkowski space-time. Since every STM atom now has an associated Minkowski space-time, it is assumed that together these   individual Minkowski space-times are equivalent to a global Minkowski operator space-time. Thus we have quantum fields on an operator  Minkowski 
space-time. In analogy with the Stueckelberg many-particle approach to relativistic quantum field theory, we equivalently describe this system as relativistic quantum mechanics on an operator space-time. From this point on, the discussion proceeds as in our earlier work \cite{Singh:2018, Singh:2019}. 

It is argued that the suppression of the operator nature of coordinate time facilitates the transition from Level I to Level II. At the same time, the operator nature of time is responsible for the novel phenomenon of quantum interference in/of time, for which we have argued that there is experimental evidence \cite{Singh:2019, Horwitz:2006, L2005}. The operator nature of spatial coordinates at Level I is interchanged for the operator nature of the canonical variables of quantum fields at Level II.This is the justification for the conventional quantisation conditions when one transits from Level III to Level II. When the operator nature of time is suppressed, Connes time is interchanged for coordinate time, in going from Level I to Level II.

The transition from level II to Level III takes place through  a GRW-type non-relativisitic spontaneous localisation \cite{Singh:2018}. In particular such localisation solves the quantum measurement problem. 

\section{Discussion}
In this work we have argued that gravity is a classical but non-commutative string theory at the Planck scale, and is described as a non-commutative geometry/matrix dynamics. The fundamental entities are `atoms' of space-time-matter in the Hilbert space -- there is no space-time, and interaction is described by the entanglement of STM atoms. Quantum general relativity emerges as the equilibrium thermodynamic approximation of the statistical thermodynamics of a large number of STM atoms.  When the statistical fluctuations about equilibrium are significant, spontaneous localisation results in the emergence of classical space-time and classical general relativity.

Mention should be made of various other approaches which derive quantum theory in some thermodynamic limit of an underlying (stochastic/dissipative/matrix/..) classical dynamics. There of course is Adler's theory of Trace Dynamics \cite{Adler:04} which is one of the prime driving forces for the present proposal. There is also the work of 't Hooft which starts from a `cellular automaton' as a fundamental unit and develops quantum theory in an emergent sense \cite{thooft}. Mention must also be made of the work of Nelson on deriving quantum mechanics from a stochastic classical dynamics \cite{Nelson}. The central difference between these works and our present proposal is that we derive not only quantum theory from an underlying dynamics but we also derive space-time geometry. To the best of our knowledge, ours is the first theory to propose a mathematically well-defined concept of a space-time-matter atom.

A number of important mathematical details remain to be worked out in this program, but the overall structure of the theory is robust and promising. In our opinion, we have proposed a new candidate quantum theory of gravity which is falsifiable. 
Ongoing experimental tests of the non-relativistic theory of spontaneous localisation \cite{RMP:2012}, which solves the quantum measurement problem, are also tests of the present quantum gravity proposal. This is because as we have argued earlier, if spontaneous localisation is the correct solution to the measurement problem, then spontaneous collapse is also necessarily responsible for the emergence of classical space-time \cite{Singh:2018}. Furthermore, our work predicts quantum interference in time, for which there possibly already is evidence \cite{L2005}, and a prediction which can be tested further. We also plan to propose an opto-mechanical experiment to look for quantum entanglement of the gravitational field of STM atoms \cite{Ulbricht:2019}.

We list below the physical and mathematical investigations which remain to be done for the completion of this program, and we very much hope that other researchers too will find these worth pursuing:

Section II: To work out in detail the low energy limit of the string theory, and the ADM formalism for this torsion gravity.

Section III: To establish the Connes form of the non-commutative torsion gravity theory, and its formulation as a matrix dynamics.

Section IV: To work out in detail the statistical thermodynamics of this matrix dynamics, and also derive the CSL theory resulting from adding stochastic corrections. To develop the relativistic spontaneous localisation model for a quantum field theory.

Section V: To work out in detail the emergence of classical space-time and classical torsion gravity, as a consequence of spontaneous localisation.

Moreover, the following observations related to this theory should also be worth investigating more closely:
\smallskip

\noindent {\it \bf A possible quantum-classical duality:} 
In the action (\ref{fundaction2}) for the STM atom,  we call an atom with string tensions $(L_1', L_2')$ `dual' to the atom with parameters  $(L_1, L_2)$ if $L_1'=L_2, L_2'=L_1)$. That is, the coupling constants for the gravity part and the torsion part have been interchanged, and one is asking if the second atom will have the same curvature as the first. If so, this suggests that corresponding to every gravity dominated STM atom, there is a dual STM atom which is torsion dominated.  Since gravity is expected to dominate for macroscopic systems and torsion for microscopic ones, this would also be a quantum-classical duality. In the classical world, the symmetric part of the metric dominates, and the anti-symmetric part dominates in the quantum world. 

\smallskip

\noindent{\bf The cosmological constant, and the expansion of the universe:} It is not clear to us at this stage whether a cosmological constant term of the form $\Lambda \Phi_{ik}$ should appear in the fundamental equations. If it does arise, then the corresponding term in the action principle should be proportional to $\Lambda \int \sqrt{\Phi} \; d^4 x$ where $\Phi$ is the determinant of the asymmetric metric. Such a term implies that $\Lambda$ is a Lagrange multiplier brought in to constrain the evolution so as to preserve four volume during evolution. However, the four-volume now depends on the anti-symmetric part of the metric as well. If spontaneous localisation is thought of as reducing the volume of the matter part of the STM atom, then to preserve the total four volume, the gravity part of the volume should increase, which might be a possible explanation for the expansion of the universe. 
It is intriguing that the proposed rate of spontaneous collapse in the GRW/CSL theory, and the rate of expansion of the universe, are comparable, both being of the order $10^{-17} \; {\rm s}^{-1}$.
Since growth of structures in the recent history of the universe implies enhanced spontaneous localisation, that might explain why the expansion is accelerating during the same epoch when structure formation and spontaneous localisation is more intense. Some attempt has been made in \cite{Singh2008} to estimate the value of the cosmological constant, using a quantum-classical duality, but more work remains to be done. Furthermore, the field equations of non-commutative torsion gravity ought to shed light on the initial conditions of the universe, and this would be worth investigating further.

\smallskip

\noindent{\bf Black hole entropy and evaporation:} In the absence of non-gravitational interactions, spontaneous localisation should result in the formation of black holes, because there is nothing to prevent the collapse from proceeding all the way to black hole formation. Thus, considering that spontaneous collapse results because of a non-unitary fluctuation away from the thermodynamic equilibrium of level I, a black hole maybe regarded as a far from equilibrium state. The process of black hole evaporation via Hawking radiation is an attempt to
annul the spontaneous fluctuation, and return to thermodynamic equilibrium. It would be interesting to study the inter-play of spontaneous collapse and Hawking evaporation, because multiple black holes of varying number of degrees of freedom could form from spontaneous localisation, during the Hawking evaporation of the original black hole. Also, the states of the  STM atom are the fundamental microstates. The entanglement of these microstates amongst various atoms might guide us towards a calculation of the entropy of the black hole which results from localisation away from equilibrium at Level I. Given the emergent  thermodynamic structure at Level I, the black hole is inevitably a thermodynamic object. Moreover, the fluctuations that result in the formation of the black hole are non-unitary, and they are present even when the black hole is evaporating through Hawking radiation. Since the entire process of black hole formation as well as evaporation has a non-unitary component, it seems to us there is no case for an information loss paradox.

\smallskip

\noindent{\bf Quantum non-locality:} Consider the entangled state of the kind shown in Eqn. (\ref{fundent}) at Level 0. This same state, when viewed on the hybrid Level II, could represent a singlet state of two quantum-correlated spins  1 and 2, far apart from each other, which are the entangled state as shown in (\ref{fundent}). On the emergent Level II, it would appear that the two spins are extremely far from each other, but note that, these same spins, when viewed at the most fundamental level, i.e. Level 0, know no notion of distance. At Level 0, we cannot say that the two spins are very far from each other - this statement is meaningless at Level 0. Distance and separation are classical emergent concepts defined only at Levels II and III. These concepts are not required for  defining an entangled state. Now when one of the particles, say 1, encounters a detector, labelled say A, it creates an entangled state of the form 
\begin{equation}
 \ket{\Psi_{\rm entangledA}} = \ket{Aa} \ket{1a} \ket{2a} +  \ket{Ab} \ket{1b} \ket{2b}
 \label{entA}
\end{equation} 
This state can be viewed from two vantage points, that of Level II, and that of Level 0. From both vantage points, the state is extremely unstable, since it is macroscopic, and undergoes rapid spontaneous collapse to $ \ket{Aa} \ket{1a} \ket{2a} $ or to $\ket{Ab} \ket{1b} \ket{2b}$. For an observer at Level II, there arises the vexing question as to how the state of particle 2 instantaneously collapses to 2b [even though it is space-like separated from 1] at the same instant at which the measurement on 1 causes its state to collapse to 1a. From the vantage point of Level 0, there is no puzzle. At Level 0, the state (\ref{entA}) just {\it is}; there is no such thing as 2 being space-like separated from 1. Detection by A simply causes spontaneous collapse to one of the two parts of the entangled state.

\smallskip

\noindent{\bf Quantum time versus classical time:} Associated with every STM atom is a quantum time observable, which is different from Connes time, and from the emergent classical time of Level III. A quantum particle (i.e. the STM atom) has a non-zero amplitude to be at more than one quantum time, at a given classical time. This is what is responsible for quantum interference in time, and for spontaneous localisation in time. Such a property could have far-reaching implications for cosmology. Time as a quantum observable in cosmology  has been recently proposed also by Magueijo and Smolin
\cite{MSm}. 

\smallskip

\noindent{\bf Norm preservation and stochastic modifications:} The evolution of the wave function, both at Level I and at Level 0, is geodesic evolution of the strings in non-commutative gravity. Such geodesic evolution preserves norm of the state, which is why norm is preserved in spite of adding non-unitary fluctuations. Thus the Born rule is possibly a consequence of geodesic motion of the state in non-commutative geometry. Moreover, the imaginary stochastic component of the metric proposed by Adler as the objective CSL noise possibly results from the stochasticity of the (away from equilibrium) correction to the anti-self adjoint antisymmetric part of the operator valued asymmetric metric. One should also investigate if the Schrodinger-Newton equation is the non-relativistic limit of the field equations of Levels 0 and III, and whether one can make contact with a recent proposal \cite{Diosi2019} on non-relativistic Planck length related quantum gravity effects.
Also, evolution at Level 0 is in general non-unitary, yet it is norm preserving because the motion is geodesic. This likely explains the non-unitary yet norm-preserving evolution in collapse models, at Levels I and II, because these models emerge from Level 0.

\smallskip

\noindent{\bf The physical meaning of the wave function:} The wave function and the particle it describes, both live in Hilbert space. Three space and four space-time are not separate or distinct from Hilbert space. Rather, space-time is what emerges from the collapse of wave function of macroscopic objects in the Hilbert space. This makes it possible to understand what physically happens with the wave function, in, say, the double slit interference experiment \cite{QTST:2017}.

\smallskip

\noindent{\bf Torsion field:} The antisymmetric torsion quantum field coming from the asymmetric part of the metric tensor is a new prediction of our theory, at the microscopic Levels 0, I and II (the field is negligible at the classical Level III). Properties of this field should be studied, so that one could look for it in suitable experiments.

\bigskip

{\bf Acknowledgements:} I am grateful to Abhinav Varma for explaining non-commutative geometry to me, and for preparing Section IIIA.  I thank Shounak De for very helpful discussions, and for help in preparing Section II. I would like to thank Angelo Bassi, Claus Kiefer, Kinjalk Lochan, T. Padmanabhan and Hendrik Ulbricht for valuable discussions over the years which have helped developing these ideas. Thanks are also due to Priyanka Giri, Navya Gupta, Manish, Shlok Nahar and Branislav Nikolic for helpful conversations. I acknowledge past grants from the John Templeton Foundation and the Foundational Questions Institute which helped lay the foundation of this programme.

\bibliography{biblioqmtstorsion}

\newpage

{\bf Response to the report of the reviewer, and List of Changes:}

Reviewer comment: This paper is a continuation of earlier papers by the same author,
with the aim to proposing an alternative quantum theory of gravity
that is also able to provide a solution to the measurement problem.
The idea is to start from a non-commutative classical spacetime and
to end up with ``atoms of space-time'' that give rise to quantum general
relativity in the thermodynamic limit.

Although this proposal is speculative and not yet presented in
a complete form, it is in my opinion interesting enough to
warrant publication. But before I can recommend doing so, the author
may wish to address the following comments.

In his Introduction, the author gives the impression that
standard approaches refer to classical time, in contrast to the
author's suggestion. But this is not the case. As it is clear from e.g.
Sec. 43.1 in Misner et al, Gravitation, the dynamics of general relativity
is the three-geometry, and there is no space-time in quantum gravity
(see in particular p. 1182 bottom and Box 43.1). This is also the case
in loop quantum gravity. The author should add some clarifying
words about this.

{\bf Author Response: I agree. The following remark has been added as a footnote on Page 2:

{It should be clarified that there is as such no space-time in quantum general relativity. The dynamical degrees of freedom are the three-geometries, or, in the case of loop quantum gravity, these are the holonomies}.

[Footnote appears as Ref 6 in References].}

Reviewer comment: The concepts of Hilbert space and quantum entanglement seem to be
put in by hand, see p. 7 and 8, in particular Eq. (5).
Is this correct? If yes, the author should state this explicity.
If no, he should point out where these concepts can be derived from.

{\bf Author Response: These concepts have been put in by hand. Hence, the following remarks have been added on p. 8:

 It should be noted that at Level 0, the concepts of Hilbert space and entanglement have been introduced by hand. These are amongst the important assumptions of the theory at Level 0.}

Reviewer comment: The fundamental entities seem to be space-time atoms, and entanglement comes from their interaction. Usually, quantum theory is not formulated on space-time, but on configuration space. In general relativity, this is the space of three-geometries (see again Misner et al, op. cit.) or the space of holonomies (in loop quantum gravity). How exactly does configuration space emerge here?

{\bf Author Response: Thank you. The following paragraph has been added to page 7:

We should also remark on the nature of configuration space for each STM atom, at Level 0. We first recall the classical torsion-gravity at Level III. Here, for the matter part, which is a closed string, the configuration variables are the string position variables. For the gravity part, in the spirit of the ADM formalism, these are the three-geometries (now built from the asymmetric metric). In going to Level 0, where non-commutativity is introduced, these same continue to be the configuration variables, except that they are now matrix-valued. Non-commutativity raises their status to that of operators / matrices. Thus the configuration variables are analogous to those in quantum general relativity, the difference now being that the commutation relations amongst them are arbitrary, and not those of quantum theory. These variables evolve in Connes time.}

Reviewer comment: On p. 24, the author states that his theory is falsifiable. How?

{\bf Author Response: Thank you. The following paragraph has been added on p. 24

Ongoing experimental tests of the non-relativistic theory of spontaneous localisation \cite{RMP:2012}, which solves the quantum measurement problem, are also tests of the present quantum gravity proposal. This is because as we have argued earlier, if spontaneous localisation is the correct solution to the measurement problem, then spontaneous collapse is also necessarily responsible for the emergence of classical space-time \cite{Singh:2018}. Furthermore, our work predicts quantum interference in time, for which there possibly already is evidence \cite{L2005}, and a prediction which can be tested further. We also plan to propose an opto-mechanical experiment to look for quantum entanglement of the gravitational field of STM atoms \cite{Ulbricht:2019}.}

Reviewer comment: The author may wish to compare his approach with other approaches
which claim to derive quantum theory in some thermodynamic limit.
For example, the cellular automata devised by G. 't Hooft (see e.g. his book ``The Cellular Automaton Interpretation of Quantum Mechanics'', Springer 2016) seem to have some resemblance to the space-time atoms, although that author does not use non-commutative geometry.

{\bf Author Response: The following paragraph has been added on p.24

Mention should be made of various other approaches which derive quantum theory in some thermodynamic limit of an underlying (stochastic/dissipative/matrix/..) classical dynamics. There of course is Adler's theory of Trace Dynamics \cite{Adler:04} which is one of the prime driving forces for the present proposal. There is also the work of 't Hooft which starts from a `cellular automaton' as a fundamental unit and develops quantum theory in an emergent sense \cite{thooft}. Mention must also be made of the work of Nelson on deriving quantum mechanics from a stochastic classical dynamics \cite{Nelson}. The central difference between these works and our present proposal is that we derive not only quantum theory from an underlying dynamics but we also derive space-time geometry. To the best of our knowledge, ours is the first theory to propose a mathematically well-defined concept of a space-time-matter atom.
}

Reviewer comment: In Eq. (4) there appears a constant with the dimension of an action.
What is its interpretation? It seems to be the same as in Eq. (6),
but I do not see this. Its expectation value (7) is a matrix, which
would not fit into (4). And what is the i in Eq. (7)? Is it the imaginary
unit? If yes, I would also not understand its occurrence in (4).

{\bf Apologies for this error. The $\tilde{C}$ in Eq. 4 has now been replaced by $\tilde{C}_0$, which is a real positive constant, which is identified with Planck?s constant below Eq. 7.}

Reviewer comment: Some minor things: Poincar\'e is misspelled on p. 10.
The author Fr\"ohlich is misspelled in Ref. [23]. Journal names in
the references should either always be abbreviated or always be written out.

{\bf Author Response: Thank you. These have been corrected.

Author Note:  Refs. 25, 35, 36 have been newly added.}

\end{document}